\documentclass[12pt]{article}
\newcommand{\pa}{\partial}
\newcommand{\A}{\alpha}
\newcommand{\B}{\beta}

\newcommand{\G}{\gamma}
\newcommand{\vp}{\varphi}

\newcommand{\ep}{\epsilon}
\newcommand{\CR}{\nonumber \\}
\newcommand{\tB}{\tilde{\beta}}
\newcommand{\tG}{\tilde{\gamma}}

\begin{document}
\begin{titlepage}
\begin{flushright}
hep-th/9910047 \\
YITP-99-59 \\
October, 1999
\end{flushright}
\vspace{0.5cm}
\begin{center}
{\Large \bf 
Green-Schwarz Superstrings on $AdS_{3}$ and the Boundary 
$N=4$ Superconformal Algebra 
}
\lineskip .75em
\vskip2.5cm
{\large Katsushi Ito}
\vskip 1.5em
{\large\it Yukawa Institute
for Theoretical Physics \\  Kyoto University, Kyoto 606-8502, Japan}  
\vskip 3.5em
\end{center}
\vskip3cm
\begin{abstract}
We study the hybrid formulation of Green-Schwarz superstrings on
$AdS_{3}$  with NS flux and the boundary $N=4$ superconformal algebra.
We show the equivalence between the NSR and GS superstrings by 
a field redefinition. The boundary $N=4$ superconformal algebra is 
realized by the free fields of the affine Lie superalgebra $A(1|1)^{(1)}$. 
We also consider the light-cone gauge and obtain the $N=4$
super-Liouville theory which describes the effective theory of
the single long string near the singularities of the D1-D5 system.

\end{abstract}
\end{titlepage}
\baselineskip=0.7cm
\newpage
The duality between type IIB superstrings on $AdS_{3}\times
S^{3}\times M^{4}$ ($M^{4}=K3$ or $T^{4}$) and two-dimensional 
$N=4$ superconformal field theory on the symmetric product of $M^{4}$ 
is an important example of the $AdS/CFT$ duality\cite{Ma}. 
It also provides the microscopic description of the black hole entropy 
\cite{MaSt}. 
The Virasoro symmetry on the boundary of $AdS_{3}$, which was firstly 
discovered by Brown and Henneaux \cite{BrHe}, can be examined by using various 
methods such as the Liouville theory obtained from the $SL(2,R)$
Chern-Simons gravity with constraints \cite{CS} and the non-linear sigma models
related to the D1-D5 system\cite{deBo}. In particular,
Giveon Kutasov and Seiberg constructed the boundary Virasoro algebra
{}from the conserved charges in the Neveu-Schwarz-Ramond (NSR)
superstring theory on $AdS_{3}\times S^{3}\times T^{4}$\cite{GKS}
(see also \cite{It,HoSu,YuZh,ESA}). 

In this paper, we study the $N=4$ superconformal
algebra on the boundary of $AdS_{3}$ from the viewpoint of the
covariantly quantized Green-Schwarz (GS) superstring theory.
The classical GS action on ten-dimensional flat space \cite{GS} or 
$AdS$ spaces\cite{AdSGS} is quite difficult to quantize due to its 
non-linearity.
On the other hand, the new (or hybrid) formulation of 
the GS superstrings compactified
on $K3$ manifolds or Calabi-Yau three-folds, proposed by 
Berkovitz and Vafa (BV) \cite{BeVa},  
have manifest Lorentz covariance and is easy to quantize since
the GS-like variables are expressed in terms of free fields.
In a recent paper\cite{BeVaWi}, Berkovitz, Vafa and Witten constructed the
GS action in both the NS and RR backgrounds, which is based on
the non-linear sigma model on the Lie supergroup $SU(2|2)'$.

One interesting property of the BV approach is that the
GS action is equivalent to that of the  covariant NSR  formalism by 
a field redefinition.
We shall construct the explicit operator map from the covariant
NSR formalism to the GS formalism. 
The action is shown to contain the affine Lie superalgebra 
$A(1|1)^{(1)}(=sl(2|2)^{(1)})$. 
The superconformal symmetry on the boundary of $AdS_{3}$ is realized 
by the Wakimoto realization of the current algebra which has been
constructed in the previous paper\cite{It}.

As an  application of this spacetime boundary 
superconformal algebra, we study the effective theory of a
long string separated from the D1-D5 system.
Seiberg and Witten\cite{SeWi} studied this system by taking the light-cone 
gauge\cite{YuZh} in the NSR formalism on $AdS_{3}$
and showed that the bosonic sector of this system becomes the 
Liouville theory.
We will show that the GS action leads to the $N=4$
superconformal Liouville theory.
This theory would be useful
to investigate the effective theory describing the 
small instanton singularity of the corresponding D1-D5 system.
 
We begin with the NSR superstrings on $AdS_{3}\times S^{3}\times
M^{4}$. Here $M^{4}$ is $K3$ or $T^{4}$. 
We then consider the correspondence between NSR and GS superstrings 
by a field redefinition.
Let us consider the $M^{4}=T^{4}$ case.
The type IIB superstrings on $AdS_{3}\times S^{3}\times T^{4}$ in the 
pure NS background is described in terms of $N=1$ supersymmetric 
$SL(2,R)$, $SU(2)$ and $U(1)^{4}$   WZW models with level $k$, which 
correspond to  the
$AdS_{3}$, $S^{3}$ and $T^{4}$ part respectively.  
We introduce fermionic ghosts $(b,c)$ with conformal 
weights $(2,-1)$ and bosonic ghosts 
$(\beta,\gamma)$ with weights $(3/2,-1/2)$.

Let $J^{\pm}(z)$, $J^{3}(z)$ be $SL(2,{\bf R})$ currents at level $k'=k+2$ 
and $K^{\pm}(z)$ , $K^{3}(z)$ the $SU(2)$ currents at level $k''=k-2$.
$\psi^{a}(z)$ and $\chi^{a}(z)$ ($a=\pm ,3$) are free fermions, which
are the superpartner of the currents $J^{a}(z)$ and $K^{a}(z)$, respectively.
Their operator product expansions (OPEs) are given by
\begin{eqnarray}
J^{+}(z)J^{-}(w)&&= {-k'\over (z-w)^{2}}+{J^{3}(w)\over z-w}+\cdots, \CR
J^{3}(z)J^{\pm}(w)&&={\pm 2 J^{\pm}(w)\over z-w}+\cdots, \quad 
J^{3}(z)J^{3}(w)={-2k'\over (z-w)^{2}}+\cdots , \\
K^{+}(z)K^{-}(w)&&= {k''\over (z-w)^{2}}+{K^{3}(w)\over z-w}+\cdots, \CR
K^{3}(z)K^{\pm}(w)&&={\pm 2 K^{\pm}(w)\over z-w}+\cdots, \quad
K^{3}(z)K^{3}(w)={2k''\over (z-w)^{2}}+\cdots , \\
\psi^{+}(z)\psi^{-}(w)&=&{1\over z-w}+\cdots, \quad
\psi^{3}(z)\psi^{3}(w)= {1\over z-w}+\cdots, \\
\chi^{+}(z)\chi^{-}(w)&=&{1\over z-w}+\cdots, \quad
\chi^{3}(z)\chi^{3}(w)= {1\over z-w}+\cdots . 
\end{eqnarray}
The currents 
\begin{eqnarray}
\hat{J}^{+}&=&J^{+}+i\sqrt{2} \psi^{+}\psi^{3}, \quad
\hat{J}^{-}=J^{-}-i\sqrt{2} \psi^{-}\psi^{3}, \quad
\hat{J}^{3}=J^{3}+2\psi^{+}\psi^{-}, \CR
\hat{K}^{+}&=&K^{+}+i\sqrt{2} \chi^{+}\chi^{3}, \quad
\hat{K}^{-}=K^{-}-i\sqrt{2} \chi^{-}\chi^{3}, \quad
\hat{K}^{3}=K^{3}+2\chi^{+}\chi^{-}, 
\label{eq:hatc}
\end{eqnarray}
obey the $SL(2,{\bf R})$ and $SU(2)$ algebra at level $k$, respectively.
The $U(1)^{4}$ current algebras are generated by the currents $i\pa Y^{j}(z)$
($j=1,2,3,4$) and the free fermions $\lambda^{j}(z)$ with the OPEs:
\begin{equation}
i \pa Y^{j}(z) i \pa Y^{k}(w)={\delta^{j k}\over (z-w)^{2}}+\ldots , \quad
\lambda^{i}(z)\lambda^{j}(w)={\delta^{ij}\over z-w}+\cdots .
\end{equation}
Now we consider the spacetime supersymmetry in the NSR formalism\cite{FMS}.
To construct the spacetime supercharges, it is convenient to bosonize
the fermions:
\begin{eqnarray}
\psi^{\pm}(z)&=& e^{\pm i H_{1}}(z), \quad 
\chi^{\pm}(z)= e^{\pm i H_{2}}(z), \quad
{1\over \sqrt{2}}(\psi_{3}\pm i\chi_{3})= e^{\pm i H_{3}}(z), \CR
{1\over \sqrt{2}}(\lambda_{1}\pm i\lambda_{2})&=& e^{\pm i H_{4}}(z), \quad
{1\over \sqrt{2}}(\lambda_{3}\pm i\lambda_{4})= e^{\pm i H_{5}}(z),
\end{eqnarray}
where $H_{I}(z)$ ($I=1,\cdots, 5$) are free bosons with OPEs
$
H_{I}(z)H_{J}(w)=-\delta_{IJ}\ln(z-w)+\cdots .
$
We also bosonize the ghost system $(b,c)$ and ($\B,\G)$:
\begin{equation}
b=e^{i\sigma},\quad c=e^{-i\sigma}, \quad
\B=-e^{-\phi}\pa\xi, \quad \G=e^{\phi}\eta,
\end{equation}
with OPEs $\sigma(z)\sigma(w)=-\ln (z-w)+\cdots$,
$\phi(z)\phi(w)=-\ln (z-w)+\cdots$ and
$\eta(z)\xi(z)=1/(z-w)+\cdots$.
The spacetime supercurrents in the $(-1/2)$-picture take the form
\begin{equation}
p_{\ep}(z)=e^{-\phi/2} S_{\epsilon}(z)
\end{equation}
where
\begin{equation}
S_{\epsilon}(z)=e^{{i\over2}\sum_{I=1}^{5}\epsilon_{I} H_{I}(z)},
\end{equation}
and $\epsilon_{I}=\pm 1$.
The $\ep_{1}, \ldots, \ep_{5}$ obey the conditions
$\prod_{I=1}^{5}\epsilon_{I}=1$ and $\prod_{I=1}^{3}\epsilon_{I}=1$ \cite{GKS}.
The supercharges are characterized by the $SL(2,{\bf
R})\times SU(2) \times U(1)$ indices $(\ep_{1},\ep_{2},\ep_{4})$.
$\epsilon_{3}$ and $\epsilon_{5}$ are 
determined by $\ep_{3}=\ep_{1}\ep_{2}$ and $\ep_{5}=\ep_{4}$. 

We discuss the GS superstrings on $AdS_{3}$.
In the BV construction\cite{BeVa}, only half of the spacetime supersymmetries
is realized in the $(-1/2)$-picture. 
These supercharges $Q_{-1/2\ \ep}$ are written in the form 
$\int dz p_{\ep}(z)$, where $p_{\ep}(z)$ are fermionic currents with
conformal weight one:
\begin{eqnarray}
p_{+++++}(z)&=& e^{-\phi/2+i/2 (H_{1}+H_{2}+H_{3})+i/2 (H_{4}+H_{5})}, \CR
p_{-+-++}(z)&=& e^{-\phi/2+i/2 (-H_{1}+H_{2}-H_{3})+i/2 (H_{4}+H_{5})}, \CR
p_{+--++}(z)&=& e^{-\phi/2+i/2 (H_{1}-H_{2}-H_{3})+i/2 (H_{4}+H_{5})}, \CR
p_{--+++}(z)&=& e^{-\phi/2+i/2 (-H_{1}-H_{2}+H_{3})+i/2 (H_{4}+H_{5})}.
\label{eq:pep}
\end{eqnarray}
Let us introduce dual variables 
$\theta^{\ep}(z)$ to $p_{\ep}(z)$ in (\ref{eq:pep}):
\begin{eqnarray}
\theta^{+++++}(z)&=& e^{\phi/2-i/2 (H_{1}+H_{2}+H_{3})-i/2 (H_{4}+H_{5})}, \CR
\theta^{-+-++}(z)&=& e^{\phi/2-i/2 (-H_{1}+H_{2}-H_{3})-i/2 (H_{4}+H_{5})}, \CR
\theta^{+--++}(z)&=& e^{\phi/2-i/2 (H_{1}-H_{2}-H_{3})-i/2 (H_{4}+H_{5})}, \CR
\theta^{--+++}(z)&=& e^{\phi/2-i/2 (-H_{1}-H_{2}+H_{3})-i/2 (H_{4}+H_{5})}.
\label{eq:tep}
\end{eqnarray}
They satisfy
\begin{equation}
p_{\ep}(z)\theta^{\ep'}(w)={\delta_{\ep}^{\ep'}\over z-w}+\cdots. 
\quad (\delta_{\ep}^{\ep'}\equiv \prod_{I=1}^{5}\delta_{\ep_{I}}^{\ep'_{I}})
\end{equation}
The pairs of fields ($p_{\ep}(z),\theta^{\ep}(z)$) are fermionic 
system with weight ($1,0$), which are fundamental fields in the BV
construction. 
In order to obtain the
full system of the GS superstrings, we need other three scalars,
which are independent of $(p_{\epsilon},\theta^{\epsilon})$.
These are given by
\begin{eqnarray}
\rho&=& -2\phi-i\kappa+i (H_{4}+H_{5}), \CR
\phi'&=& {1\over\sqrt{2}} (H_{4}-H_{5}), \CR
\phi''&=& {1\over \sqrt{2}} (H_{4}+H_{5}-2\kappa+2i\phi),
\end{eqnarray}
where we have introduced a scalar field $\kappa(z)$ by
$(\eta,\xi)=(e^{i\kappa}, e^{-i\kappa})$.
Since the $U(1)$ current $i\pa (H_{4}+H_{5})(z)$ of $T^{4}$ is changed to 
$\sqrt{2}i\pa\phi''(z)=i\pa (H_{4}+H_{5})(z)-2\pa (\phi+i\kappa)(z)$,
the free fermions $\lambda_{i}$ must be replaced by $\hat{\lambda}_{i}$, 
where
\begin{equation}
{1\over \sqrt{2}}(\hat{\lambda}_{1}\pm i\hat{\lambda}_{2})=
 e^{\pm i (H_{4}-\kappa+i\phi)}(z), \quad
{1\over \sqrt{2}}(\hat{\lambda}_{3}\pm i\hat{\lambda}_{4})=
 e^{\pm i (H_{5}-\kappa+i\phi)}(z).
\end{equation}
To summarize, the $SL(2,R)$ and $SU(2)$ currents
$J^{a}(z)$, $K^{a}(z)$ ($a=\pm,3$), the fermionic ghosts 
$(p_{\ep}(z),\theta^{\ep}(z))$, two free bosons $\sigma(z)$,
$\rho(z)$, $N=1$ $U(1)^{4}$ superconformal fields
$(Y^{i}(z),\hat{\lambda}_{i}(z))$ are  fundamental fields in the GS action on
$AdS_{3}\times S^{3}\times T^{4}$.

For $M^{4}=K3$ case, we may use the $N=4$ superconfomal field theory with
central charge $c=6$ instead of $(Y^{i},\lambda_{i})$. 
We take an $N=2$ superconformal subalgebra
$(T_{K3},G^{\pm}_{K3},J_{K3})$ by choosing the $U(1)$
current $J_{K3}(z)=i\pa H_{K3}(z)$ where
\begin{equation}
J_{K3}(z)J_{K3}(w)={\hat{c} \over (z-w)^{2}}+\cdots,
\end{equation}
with $\hat{c}=c/3=2$.
The spin operators $(p_{\epsilon},\theta^{\epsilon})$ and the scalar
field $\rho$ are obtained by 
replacing $H_{4}+H_{5}$ by $H_{K3}$. 
The $U(1)$ field is shifted by the $-2\pa(\phi+i\kappa)$ so that the 
supercurrents $G^{\pm}_{K3}$ are modified as
$\hat{G}^{\pm}_{K3}=e^{\pm(\phi+i\kappa)} G^{\pm}_{K3}$.

We now discuss the spacetime superconformal symmetry on the boundary
of $AdS_{3}$. 
Half of the spacetime supercharges are realized by the (-1/2)-picture
supercharges $Q_{-1/2\ \ep}$.
Other supercharges are realized in the $(+1/2)$-picture. 
Let us denote them as $Q_{+1/2}^{\ep}$, which are 
expressed in terms of ($p_{\ep}(z),\theta^{\ep}(z)$)\cite{BeVaWi}:
\begin{equation}
Q_{+1/2}^{\ep}=-i \int dz (i p_{\ep} e^{-\sigma-i\rho}+ q^{\ep}),
\end{equation}
where
\begin{eqnarray}
q^{+++++}(z)&=& {1\over2}\theta_{+++++}(\hat{J}^{3}+\hat{K}^{3})
-\theta^{-+-++}K^{-}-\theta^{+--++}J^{-} \CR
&& -\theta^{-+-++}\theta^{+--++}p_{--+++}+k\pa\theta^{+++++},
\CR
q^{-+-++}(z)&=& {1\over2}\theta_{-+-++}(-\hat{J}^{3}+\hat{K}^{3})
-\theta^{--+++}K^{+}-\theta^{+++++}J^{-} \CR
&& -\theta^{--+++}\theta^{+++++}p_{-+-++}+k\pa\theta^{-+-++},
\CR
q^{+--++}(z)&=& {1\over2}\theta_{+--++}(\hat{J}^{3}-\hat{K}^{3})
-\theta^{+++++}K^{-}-\theta^{--+++}J^{+} \CR
&& -\theta^{+++++}\theta^{--+++}p_{+--++}+k\pa\theta^{+--++},
\CR
q^{--+++}(z)&=&-{1\over2}\theta_{--+++}(\hat{J}^{3}+\hat{K}^{3})
-\theta^{+--++}K^{+}-\theta^{-+-++}J^{+} \CR
&& -\theta^{-+-++}\theta^{+--++}p_{+++++}+k\pa\theta^{--+++}.
\label{eq:sup2}
\end{eqnarray}
Here  the $SL(2,{\bf R})$ and $SU(2)$ 
currents $\hat{K}$ and $\hat{J}$ in (\ref{eq:hatc}) 
are expressed in terms of 
($p_{\ep},\theta^{\ep}$) as follows:
\begin{eqnarray}
\hat{K}^{-}&=& K^{-}+\theta^{--+++}p_{+--++}+\theta^{+--++}p_{--+++},\CR
\hat{K}^{+}&=& K^{+}+\theta^{-+-++}p_{+++++}+\theta^{+++++}p_{-+-++},\CR
\hat{K}^{3}&=& K^{0}+\theta^{-+-++}p_{-+-++}+\theta^{+++++}p_{+++++}
-\theta^{--+++}p_{--+++}-\theta^{+--++}p_{+--++}, \CR
\hat{J}^{-}&=& J^{-}-\theta^{+++++}p_{+--++}-\theta^{--+++}p_{-+-++}, \CR
\hat{J}^{+}&=& J^{+}-\theta_{+++++}p_{+--++}-\theta^{+--++}p_{+++++}, \CR
\hat{J}^{3}&=& J^{0}+\theta^{-+-++}p_{-+-++}-\theta^{+++++}p_{+++++}
+\theta^{--+++}p_{--+++}-\theta^{+--++}p_{+--++}. \CR
\end{eqnarray}
We study the subalgebra which does not depend on the
term $p_{\epsilon}e^{-\rho-i\rho}$,
although such terms play an important role to construct the
GS action in the RR background\cite{BeVaWi}.
We may show that the currents 
$p_{\ep}(z)$, $q^{\ep}(z)$, $\hat{K}^{\pm,3}(z)$, $\hat{J}^{\pm,3}(z)$
satisfy the affine Lie superalgebra $sl(2|2)^{(1)}$ at level $k$.

Let us review the structure of the Lie superalgebra $sl(2|2)$\cite{Ka}.
This algebra has the simple roots  $\A_{i}$ ($i=1,2,3$), 
where  $\A_{1}=e_{1}-e_{2}$, $\A_{2}=e_{2}-\delta_{1}$,
$\A_{3}=\delta_{1}-\delta_{2}$ and  $e_{a}$ ($\delta_{a}$) are the
orthonormal basis with positive (negative) norm: $e_{a}\cdot
e_{b}=\delta_{ab}$ ($\delta_{a}\cdot\delta_{b}=-\delta_{ab}$).
$sl(2|2)$ has rank $r=2$ since the algebra is obtained by taking the
quotient by the one-dimensional ideal generated by the identity matrix.
An affine Lie superalgebra with rank $r$ is generated by 
bosonic currents $J_{\pm\A}(z)$ ($\A\in\Delta^{+}_{0}$), the 
Cartan currents $H^{i}(z)$ ($i=1,\ldots, r$)
and fermionic currents $j_{\pm\A}(z)$ ($\A\in\Delta^{+}_{1}$). 
Here $\Delta^{+}_{0}$ ($\Delta^{+}_{1}$) denotes the set of positive
even (odd) roots.
For $sl(2|2)$, we have 
$\Delta^{+}_{0}=\{ \A_{1},\A_{3}\}$
and $\Delta^{+}_{1}=\{ \A_{2}, \A_{1}+\A_{2},\A_{2}+\A_{3}, 
\A_{1}+\A_{2}+\A_{3} \}$.
The bosonic currents $J_{\pm\A_{1}}(z)$ and $\A_{1}\cdot H(z)$
obey the $SU(2)$ affine Lie algebra at level $k$.
The currents $J_{\pm\A_{3}}(z)$ and $\A_{3}\cdot H(z)$ satisfy the 
$SL(2,{\bf R})$ affine Lie algebra at level $k$.
The pairs of fermionic currents
($j_{\pm(\A_{1}+\A_{2})}(z)$, $j_{\pm\A_{2}}(z)$) and 
($j_{\pm(\A_{1}+\A_{2}+\A_{3})}(z)$, $j_{\pm(\A_{2}+\A_{3})}(z)$)
form the doublets with respect to the $SU(2)$ subalgebra.
We may identify the currents $p_{\ep}(z)$, $q^{\ep}(z)$,
$\hat{K}^{\pm,3}(z)$, $\hat{J}^{\pm,3}(z)$
in terms of the currents of $sl(2|2)^{(1)}$ as follows:
\begin{eqnarray}
j_{-\A_{2}}(z)&=& p_{+++++}(z), \quad
j_{-\A_{1}-\A_{2}}(z)= p_{+--++}(z), \CR
j_{-\A_{2}-\A_{3}}(z)&=& p_{-+-++}(z), \quad
j_{-\A_{1}-\A_{2}-\A_{3}}(z)= p_{--+++}(z), \CR
j_{\A_{1}+\A_{2}+\A_{3}}(z)&=& q^{+++++}(z), \quad
j_{\A_{2}+\A_{3}}(z)= q^{+--++}(z), \CR
j_{\A_{1}+\A_{2}}(z)&=& q^{-+-++}(z), \quad
j_{\A_{2}}(z)= q^{--+++}(z), \CR
J_{\A_{1}}(z)&=& \hat{K}^{+}(z), \quad J_{-\A_{1}}(z)=\hat{K}^{-}(z), \quad
\A_{1}\cdot H(z)=\hat{K}^{3}(z), \CR
J_{\A_{3}}(z)&=& \hat{J}^{+}(z), \quad J_{-\A_{3}}(z)=\hat{J}^{-}(z), \quad
\A_{3}\cdot H(z)=\hat{J}^{3}(z).
\label{eq:waki}
\end{eqnarray}
When we apply the Wakimoto realization for the $SL(2,{\bf R})$ and 
$SU(2)$ currents $J^{\pm,3}(z)$ and $K^{\pm,3}(z)$, we obtain the free
field realization of the  affine Lie superalgebra $sl(2|2)^{(1)}$.
Introduce the free bosons $\vp^{i}(z)$ and two pairs of bosonic ghost
system $(\tB_{\A_{1}},\tG_{\A_{1}})$ and
$(\tB_{\A_{3}},\tG_{\A_{3}})$, 
whose conformal weights are $(1,0)$. 
Their OPEs are given by
\begin{equation}
\vp^{i}(z)\vp^{j}(w)=-\delta^{ij}\ln(z-w)+\cdots, \quad
\tB_{\A}(z)\tG_{\A'}={\delta^{\A,\A'}\over z-w}+\cdots.
\end{equation}
In terms of these free fields, the currents $J$ and $K$ are 
written as
\begin{eqnarray}
K^{-}&=& \tB_{\alpha_{1}},\CR
K^{3}&=& -i\A_{+}\A_{1}\cdot\pa\vp+2\tG_{\alpha_{1}}\tB_{\alpha_{1}},\CR
K^{+}&=&
i\A_{+}\A_{1}\cdot\pa\vp\tG_{\alpha_{1}}-\tG_{\alpha_{1}}^{2}\tB_{\alpha_{1}}
+(k-2)\pa\tG_{\alpha_{1}},
\\
J^{-}&=& \tB_{\alpha_{3}},\CR
J^{3}&=& -i\A_{+}\A_{3}\cdot\pa\vp-2\tG_{\alpha_{3}}\tB_{\alpha_{3}},\CR
J^{+}&=& -i\A_{+}\A_{3}\cdot\pa\vp\tG_{\alpha_{3}}
-\tG_{\alpha_{3}}^{2}\tB_{\alpha_{3}}-(k+2)\pa\tG_{\alpha_{3}},
\end{eqnarray}
where $\A_{+}=\sqrt{k}$. 
We also identify the $(p_{\ep},\theta^{\ep})$ system to the fermionic
ghosts $(\tilde{\eta}_{\A},\tilde{\xi}_{\A})$ ($\A\in\Delta^{+}_{1}$):
\begin{eqnarray}
\tilde{\eta}_{\A_{2}}&=& p_{+++++}, \quad 
\tilde{\eta}_{\A_{1}+\A_{2}}= p_{+--++}, \quad 
\tilde{\eta}_{\A_{2}+\A_{3}}= p_{-+-++}, \quad 
\tilde{\eta}_{\A_{1}+\A_{2}+\A_{3}}= p_{--+++}, \CR
\tilde{\xi}_{\A_{2}}&=& \theta^{+++++}, \quad 
\tilde{\xi}_{\A_{1}+\A_{2}}= \theta^{+--++}, \quad 
\tilde{\xi}_{\A_{2}+\A_{3}}= \theta^{-+-++}, \quad 
\tilde{\xi}_{\A_{1}+\A_{2}+\A_{3}}= \theta^{--+++}.
\end{eqnarray}
This free field realization is different from the one given in \cite{It}. 
We find that the relation between two realizations is given by the formulae
\begin{eqnarray}
\G_{\alpha_{1}}&=& \tG_{\alpha_{1}}, \quad
\B_{\alpha_{1}}
= \tB_{\alpha_{1}}+\tilde{\xi}_{\A_{2}}\tilde{\eta}_{\A_{2}+\A_{2}}
+\tilde{\xi}_{\A_{2}+\A_{3}}\tilde{\eta}_{\A_{1}+\A_{2}+\A_{3}}, \CR
\G_{\alpha_{3}}&=&\tG_{3}, \quad 
\B_{\alpha_{3}}= \tB_{\alpha_{3}}, \CR
\xi_{\alpha_{2}}&=& \tilde{\xi}_{\A_{2}}, \quad
\eta_{\alpha_{2}}=\tilde{\eta}_{\A_{2}}+\tG_{1}\tilde{\eta}_{\A_{1}+\A_{2}},
 \CR
\xi_{\A_{1}+\A_{2}}&=&\tilde{\eta}_{\A_{1}+\A_{2}}
-\tG_{\alpha_{1}}\tilde{\xi}_{\A_{1}+\A_{2}+\A_{3}},\quad
\eta_{\A_{1}+\A_{2}}=\tilde{\eta}_{\A_{1}+\A_{2}},\CR
\xi_{\A_{2}+\A_{3}}&=&\tilde{\xi}_{\A_{2}+\A_{3}}, \quad
\eta_{\A_{2}+\A_{3}}=\tilde{\eta}_{\A_{2}+\A_{3}}
+\tG_{\alpha_{1}}\tilde{\eta}_{\A_{1}+\A_{2}+\A_{3}}, \CR
\xi_{\A_{1}+\A_{2}+\A_{3}}&=&\tilde{\xi}_{\A_{1}+\A_{2}+\A_{3}}
-\tG_{\alpha_{1}}\tilde{\xi}_{\A_{1}+\A_{2}+\A_{3}},\quad
\eta_{\A_{1}+\A_{2}+\A_{3}}=\tilde{\eta}_{\A_{1}+\A_{2}+\A_{3}},
\end{eqnarray}
where $(\beta_{\alpha},\gamma_{\alpha})$ and
($\eta_{\alpha},\xi_{\alpha}$)
are free fields used in \cite{It}.
In the previous work\cite{It}, we have shown that  the 
free field realization of affine Lie superalgebra $sl(2|2)^{(1)}$ 
leads to the spacetime $N=4$ superconformal algebra on
the boundary of $AdS_{3}$. 
By using the operator map between the NSR and the GS
strings,  we can show that the  boundary $N=4$ superconformal
algebra constructed in the NSR 
formalism \cite{GKS,HoSu} is  equivalent to the algebra in \cite{It}.
In the present paper, we examine the spacetime boundary
superconformal algebra by taking the ``light-cone'' 
gauge\cite{YuZh,SeWi}.
In the light-cone gauge, the spacetime 
Virasoro algebra turns out to be the worldsheet one. 
The theory is expected to describe 
the effective theory of the single long string separated from
the D1/D5 system\cite{SeWi}.
We take the following the light-cone gauge 
in (\ref{eq:waki}):
\begin{eqnarray}
&& \A_{3}\cdot\vp=\vp_{0} \gg 1, \quad
\G_{\A_{3}}=z,\quad \bar{\G}_{\A_{3}}=\bar{z}, \CR
&& \tilde{\xi}_{\A_{2}+\A_{3}}=-z \tilde{\xi}_{\A_{2}},
 \quad \tilde{\xi}_{\A_{1}+\A_{2}+\A_{3}}=-z \tilde{\xi}_{\A_{1}+\A_{2}}.
\label{eq:llg}
\end{eqnarray}
Since the fermionic fields
$\tilde{\xi}_{\alpha_{1}+\alpha_{2}+\alpha_{3}}$ and 
$\tilde{\xi}_{\alpha_{2}+\alpha_{3}}$ are written in terms of other fields,
we put
$\tilde{\eta}_{\alpha_{1}+\alpha_{2}+\alpha_{3}}=0$ and 
$\tilde{\eta}_{\alpha_{2}+\alpha_{3}}=0$.
In order to study the spacetime conformal symmetry on the boundary of 
$AdS_{3}$, we need to solve the constraint $T_{total}=0$, where 
the total energy-momentum (EM)tensor is defined by 
$T_{total}=T_{sl(2|2)}+T_{M^{4}}+T_{ghosts}$.
$T_{sl(2|2)}$ denotes the EM tensor of the
$sl(2|2)$ WZW model. 
$T_{M^{4}}$ is that of the conformal field
theory on $M^{4}$ and $T_{ghosts}$ is constructed out of $(\sigma,\rho)$
bosons. 
In the classical approximation, we may ignore the ghost contribution. 
$T_{sl(2|2)}$ is then given by
\begin{eqnarray}
T_{sl(2|2)}&=&
-{1\over 2} (\partial\varphi)^{2}-{i \over 2\alpha_{+}}
 (\alpha_{1}+\alpha_{3}) \cdot \partial^{2} \varphi 
+\tilde{\beta}_{\alpha_{1}}\partial\tilde{\gamma}_{\alpha_{1}}
+\tilde{\beta}_{\alpha_{3}}\partial\tilde{\gamma}_{\alpha_{3}} \CR
&& -\tilde{\eta}_{\alpha_{2}}\partial\tilde{\xi}_{\alpha_{2}}
-\tilde{\eta}_{\alpha_{1}+\alpha_{2}}
\partial\tilde{\xi}_{\alpha_{1}+\alpha_{2}}
-\tilde{\eta}_{\alpha_{2}+\alpha_{3}}
\partial\tilde{\xi}_{\alpha_{2}+\alpha_{3}}
-\tilde{\eta}_{\alpha_{1}+\alpha_{2}+\alpha_{3}}
\partial\tilde{\xi}_{\alpha_{1}+\alpha_{2}+\alpha_{3}}.
\end{eqnarray}
Solving the constraint $T_{total}=0$ with respect to 
$\tilde{\beta}_{\alpha_{3}}$, we obtain 
$\tilde{\beta}_{\alpha_{3}}=-\tilde{T}-T_{M^{4}}$, where 
\begin{equation} 
\tilde{T}(z)=
-{1\over 2} (\partial\varphi)^{2}
-{i \over 2\alpha_{+}}
 (\alpha_{1}+\alpha_{3}) \cdot \partial^{2} \varphi 
+\tilde{\beta}_{\alpha_{1}}\partial\gamma_{\alpha_{1}}
-\tilde{\eta}_{\alpha_{2}}\partial\tilde{\xi}_{\alpha_{2}}
-\tilde{\eta}_{\alpha_{1}+\alpha_{2}}
\partial\tilde{\xi}_{\alpha_{1}+\alpha_{2}}.
\label{eq:tilt}
\end{equation}
Let us consider the spacetime Virasoro algebra in this light-cone 
gauge. 
The generator $L_{n}$ ($n\in{\bf Z}$) of the Virasoro algebra is 
expressed as $L_{n}=\int {d z\over 2\pi i} {\cal
L}_{n}(z)$\cite{GKS}, 
where
\begin{equation}
 {\cal L}_{n}(z)={1-n^2\over 2} \alpha_{3}\cdot H \gamma^{n}
-{n(n-1)\over 2}J_{-\alpha_{3}}\gamma^{n+1}
+{n(n+1)\over 2}J_{\alpha_{3}}\gamma^{n-1}.
\end{equation}
Here $\gamma=\tilde{\gamma}_{\alpha_{3}}$.
In the light-cone like gauge, using the free field realization
(\ref{eq:waki}) and $\tilde{\beta}_{\alpha_{3}}=-\tilde{T}-T_{M^{4}}$,  
${\cal L}_{n}(z)$ is shown to be
\begin{equation}
{\cal L}_{n}(z)= 
-z^{n+1}(\tilde{T}+T_{M^{4}}) +{n+1\over2} z^{n}
\left(-i\alpha_{+}\alpha_{3}\cdot\partial\varphi+
\tilde{\xi}_{\alpha_{2}}\tilde{\eta}_{\alpha_{2}}
+\tilde{\xi}_{\alpha_{1}+\alpha_{2}}\tilde{\eta}_{\alpha_{1}+\alpha_{2}}
\right). 
\label{eq:ln}
\end{equation}
{}From (\ref{eq:tilt}), it is shown that  $L_{n}$
takes of the form $\int {d z\over 2\pi i}z^{n+1}T(z)$, where 
$T(z)=T_{N=4}(z)+T_{M^{4}}(z)$ and $T_{N=4}(z)$ is the EM tensor of 
the $N=4$ $SU(2)$ extended superconformal algebra with central charge 
$c=6(k-1)$: 
\begin{eqnarray}
T_{N=4}(z)&=&
-{1\over 2} (\partial\varphi)^{2}
-{i \over 2\alpha_{+}}
 \alpha_{1}\cdot \partial^{2} \varphi 
-{i \over 2}({1\over \alpha_{+}}-\alpha_{+})\alpha_{3}\cdot\pa^{2}\varphi
+\tilde{\beta}_{\alpha_{1}}\partial\tilde{\gamma}_{\alpha_{1}} \CR
&& -{1\over2}\left(
\tilde{\eta}_{\alpha_{2}}\partial\tilde{\xi}_{\alpha_{2}}
-(\partial \tilde{\eta}_{\alpha_{2}})\tilde{\xi}_{\alpha_{2}}\right)
-{1\over2}\left(\tilde{\eta}_{\alpha_{1}+\alpha_{2}}
\partial\tilde{\xi}_{\alpha_{1}+\alpha_{2}}
-(\partial\tilde{\eta}_{\alpha_{1}+\alpha_{2}})
\tilde{\xi}_{\alpha_{1}+\alpha_{2}}
\right).
\label{eq:emnf}
\end{eqnarray}
We note that  the EM tensor (\ref{eq:emnf}) has no quantum correction 
though we treat the constraints in a classical way.
In the light-cone gauge, the fermionic ghosts
$(\tilde{\eta}_{\alpha_{1}+\alpha_{2}},
\tilde{\xi}_{\alpha_{1}+\alpha_{2}})$
and 
$(\tilde{\eta}_{\alpha_{2}},
\tilde{\xi}_{\alpha_{2}})$
become the fermions with spin $(1/2,1/2)$ due to the twist in (\ref{eq:ln}).
Adding $T_{M^{4}}$ with the central charge $c=6$, 
the worldsheet total EM tensor $T(z)$ has the central charge $c=6k$ as
expected from the spacetime Virasoro symmetry.

The spacetime $SU(2)$ affine Lie algebra symmetry generated by \cite{It}
\begin{equation}
T^{\pm }_{n}=\oint {d z\over2\pi i}
{\cal T}^{\pm}_{n}(z), \quad
T^{0}_{n}=\oint {d z\over2\pi i}
{\cal T}^{0}_{n}(z),
\end{equation}
where 
\begin{eqnarray}
{\cal T}^{+}_{n}(z)&=& J_{\A_{1}}\G^{n}+n
\widehat{\xi}_{\A_{1}+\A_{2}+\A_{3}} \widehat{\eta}_{\A_{2}}\G^{n-1}, \CR
{\cal T}^{-}_{n}(z)&=& J_{-\A_{1}}\G^{n}+n
\widehat{\xi}_{\A_{2}+\A_{3}} \widehat{\eta}_{\A_{1}+\A_{2}}\G^{n-1}, \CR
{\cal T}^{0}_{n}(z)&=& {1\over2}\A_{1}\cdot H\G^{n}
+{n\over2} \widehat{\xi}_{\A_{1}+\A_{2}+\A_{3}}
\widehat{\eta}_{\A_{1}+\A_{2}}\G^{n-1} 
-{n\over2} \widehat{\xi}_{\A_{2}+\A_{3}}
\widehat{\eta}_{\A_{2}}\G^{n-1} , 
\end{eqnarray}
and 
\begin{eqnarray}
\widehat{\xi}_{\A_{2}+\A_{3}}&=&
\tilde{\xi}_{\A_{2}+\A_{3}}+\tilde{\G}_{\A_{3}}\tilde{\xi}_{\A_{2}}, \quad
\widehat{\xi}_{\A_{1}+\A_{2}+\A_{3}}= \tilde{\xi}_{\A_{1}+\A_{2}+\A_{3}}
+\G_{\A_{3}}\tilde{\xi}_{\A_{1}+\A_{2}}, \CR
\widehat{\eta}_{\A_{1}+\A_{2}}&=&\tilde{\eta}_{\A_{1}+\A_{2}}
-\tilde{\G}_{\A_{3}}\tilde{\eta}_{\A_{1}+\A_{2}+\A_{3}}, \quad
\widehat{\eta}_{\A_{2}}= \tilde{\eta}_{\A_{2}}
-\tilde{\G}_{\A_{3}}\tilde{\eta}_{\A_{2}+\A_{3}}.
\end{eqnarray}
In the light-cone gauge (\ref{eq:llg}), we obtain
$\widehat{\xi}_{\A_{2}+\A_{3}}=\widehat{\xi}_{\A_{1}+\A_{2}+\A_{3}}=0$,
$\widehat{\eta}_{\A_{1}+\A_{2}}=\tilde{\eta}_{\A_{1}+\A_{2}}$ and
$\widehat{\eta}_{\A_{1}+\A_{2}+\alpha_{3}}
=\tilde{\eta}_{\A_{1}+\A_{2}+\alpha_{3}}$. 
Therefore $T^{\pm}_{n}$ and $T^{0}_{n}$ lead to the worldsheet $SU(2)$ 
affine Lie algebra generated by the currents 
$T^{a}(z)=\sum_{n}T_{n}^{a}z^{-n-1}$:
\begin{eqnarray}
T^{+}(z)&=& K^{+}+\tilde{\xi}_{\alpha_{1}+\alpha_{2}}
 \tilde{\eta}_{\A_{2}}, \CR
T^{-}(z)&=& K^{-}+\tilde{\xi}_{\alpha_{2}}
 \tilde{\eta}_{\alpha_{1}+\A_{2}},\CR
T^{0}(z)&=& K^{3}
+\tilde{\xi}_{\alpha_{1}+\alpha_{2}} \tilde{\eta}_{\alpha_{1}+\A_{2}}
-\tilde{\xi}_{\alpha_{2}} \tilde{\eta}_{\alpha_{2}}.
\end{eqnarray}
Now we consider the space-time superconformal symmetry in the light-cone 
gauge.
Since we have imposed the constraints
$\tilde{\eta}_{\alpha_{2}+\alpha_{3}}
=\tilde{\eta}_{\alpha_{1}+\alpha_{2}+\alpha_{3}}=0$, which break the
half of spacetime supersymmetries,  we expect to obtain half of
generators of the $N=4$ spacetime superconformal algebra.
The supercurrents of the spacetime $N=4$ superconformal algebra in the
NS sector are given by \cite{It}
\begin{equation}
G^{a}_{r}=\oint {dz\over 2\pi i} {\cal G}^{a}_{r}(z), \quad 
\bar{G}^{a}_{r}=\oint {dz\over 2\pi i} \bar{{\cal G}}^{a}_{r}(z), 
\quad (a=1,2,  \ \ r\in {\bf Z}+{1\over2})
\end{equation}
where 
\begin{eqnarray}
{\cal G}^{1}_{n-1/2}(z)&=&-(n-1) j_{-\A_{1}-\A_{2}}\G^{n} +n
j_{-\A_{1}-\A_{2}-\A_{3}} \G^{n-1}, \CR
{\cal G}^{2}_{n-1/2}(z)&=&-(n-1) j_{-\A_{2}}\G^{n} +n
j_{-\A_{2}-\A_{3}} \G^{n-1}, \CR
\bar{{\cal G}}^{1}_{n-1/2}(z)&=& -(n-1) j_{\A_{1}+\A_{2}}\G^{n} - n
j_{\A_{1}+\A_{2}+\A_{3}} \G^{n-1} \CR
&& -n (n-1) \left(
\A_{3}\cdot H \G^{n-1}\tilde{\xi}_{\A_{1}+\A_{2}+\A_{3}}
+J_{-\A_{3}}\G^{n}\tilde{\xi}_{\A_{1}+\A_{2}+\A_{3}}
-J_{\A_{3}}\G^{n-2}\tilde{\xi}_{\A_{1}+\A_{2}+\A_{3}}
\right), \CR
\bar{{\cal G}}^{2}_{n-1/2}(z)&=& -(n-1) j_{\A_{2}}\G^{n} - n
j_{\A_{2}+\A_{3}} \G^{n-1} \CR
&&
\!\!\!\!\!\!\!\!\!\!
 -n (n-1) \left(
\A_{3}\cdot H \G^{n-1}\tilde{\xi}_{\A_{2}+\A_{3}}
+J_{-\A_{3}}\G^{n}\tilde{\xi}_{\A_{2}+\A_{3}}
-J_{\A_{3}}\G^{n-2}\tilde{\xi}_{\A_{2}+\A_{3}}
\right). \quad  (n\in {\bf Z})
\end{eqnarray}
In the light-cone gauge, we obtain
\begin{eqnarray}
{\cal G}^{1}_{n-1/2}(z)&=&-(n-1) \tilde{\eta}_{\alpha_{1}+\alpha_{2}}z^{n},
\CR
{\cal G}^{2}_{n-1/2}(z)&=&-(n-1) \tilde{\eta}_{\alpha_{2}}z^{n},
\CR
\bar{{\cal G}}^{1}_{n-1/2}(z)&=& 
\left( -{1\over2}\tilde{\xi}_{\alpha_{1}+\alpha_{2}}
i\alpha_{+}\alpha_{3}\cdot\partial\varphi
-{1\over2} 
\tilde{\xi}_{\alpha_{1}+\alpha_{2}} K^{3}
-\tilde{\xi}_{\alpha_{2}} K^{+}
+\tilde{\xi}_{\alpha_{1}+\alpha_{2}} \tilde{\xi}_{\alpha_{2}}
\tilde{\eta}_{\alpha_{2}}
+k\pa\tilde{\xi}_{\alpha_{1}+\alpha_{2}} \right) z^{n} \CR
&& -2 n \tilde{\xi}_{\alpha_{1}+\alpha_{2}} z^{n-1}, \CR
\bar{{\cal G}}^{2}_{n-1/2}(z)&=& 
\left( -{1\over2}\tilde{\xi}_{\alpha_{2}}
i\alpha_{+}\alpha_{3}\cdot\partial\varphi
+{1\over2} 
\tilde{\xi}_{\alpha_{2}} K^{3}
-\tilde{\xi}_{\alpha_{1}+\alpha_{2}} K^{-}
+\tilde{\xi}_{\alpha_{2}} \tilde{\xi}_{\alpha_{1}+\alpha_{2}} 
\tilde{\eta}_{\alpha_{1}+\alpha_{2}}
+k\pa\tilde{\xi}_{\alpha_{2}} \right) z^{n} \CR
&& -2 n \tilde{\xi}_{\alpha_{2}} z^{n-1}.
\label{eq:supg}
\end{eqnarray}
The first two currents in (\ref{eq:supg}) correspond to the world-sheet
fermions. 
The last two currents, on the other hand, give rise to the supercurrents 
of the $N=4$ superconformal algebra.
Including the quantum correction, we obtain the worldsheet supercurrents
$\bar{G}^{a}(z)=\sum_{r}\bar{G}_{r}^{a}z^{-r-3/2}$:
\begin{eqnarray}
&&\!\!\!\!\!\!\!\!\!
\bar{G}^{1}(z)= {i\over2}\alpha_{+}\alpha_{3}\cdot\pa\varphi
 \tilde{\xi}_{\alpha_{1}+\alpha_{2}}
+(k-1)\partial  \tilde{\xi}_{\alpha_{1}+\alpha_{2}}
-{1\over2} \tilde{\xi}_{\alpha_{1}+\alpha_{2}}K^{3}
-\tilde{\xi}_{\alpha_{2}} K^{+}
+\tilde{\xi}_{\alpha_{1}+\alpha_{2}}
\tilde{\xi}_{\alpha_{2}}\tilde{\eta}_{\alpha_{2}}, \CR
&&\!\!\!\!\!\!\!\!\!
\bar{G}^{2}(z)= {i\over2}\alpha_{+}\alpha_{3}\cdot\pa\varphi
 \tilde{\xi}_{\alpha_{2}}+(k-1)\partial  \tilde{\xi}_{\alpha_{2}}
+{1\over2} \tilde{\xi}_{\alpha_{2}}K^{3}
-\tilde{\xi}_{\alpha_{1}+\alpha_{2}} K^{-}
+\tilde{\xi}_{\alpha_{2}}
\tilde{\xi}_{\alpha_{1}+\alpha_{2}}\tilde{\eta}_{\alpha_{1}+\alpha_{2}}.
\end{eqnarray}
Remaining worldsheet supercurrents must be introduced by hand.
These are obtained by replacing
$\tilde{\xi}$ fields by $\tilde{\eta}$:
\begin{eqnarray}
&&\!\!\!\!\!\!\!\!\!
G^{1}(z)= -{i\over2}\alpha_{+}\alpha_{3}\cdot\pa\varphi
 \tilde{\eta}_{\alpha_{1}+\alpha_{2}}
+(1-k)\partial  \tilde{\eta}_{\alpha_{1}+\alpha_{2}}
-{1\over2} \tilde{\eta}_{\alpha_{1}+\alpha_{2}}K^{3}
-\tilde{\eta}_{\alpha_{2}} K^{-}
+\tilde{\eta}_{\alpha_{1}+\alpha_{2}}
\tilde{\xi}_{\alpha_{2}}\tilde{\eta}_{\alpha_{2}}, \CR
&&\!\!\!\!\!\!\!\!\!
G^{2}(z)= -{i\over2}\alpha_{+}\alpha_{3}\cdot\pa\varphi
 \tilde{\eta}_{\alpha_{2}}+(1-k)\partial  \tilde{\eta}_{\alpha_{2}}
+{1\over2} \tilde{\eta}_{\alpha_{2}}K^{3}
-\tilde{\eta}_{\alpha_{1}+\alpha_{2}} K^{+}
+\tilde{\eta}_{\alpha_{2}}
\tilde{\xi}_{\alpha_{1}+\alpha_{2}}\tilde{\eta}_{\alpha_{1}+\alpha_{2}}.
\end{eqnarray}
It is shown that 
the operators $T_{N=4}(z)$, $G^{a}(z)$, $\bar{G}^{a}(z)$, 
and $T^{\pm,0}(z)$ generate the $N=4$ $SU(2)$
extended superconformal algebra with the central charge $c=6(k-1)$.

In the present paper, we have discussed the relation between the NSR
formulation and the GS formulation of the superstrings on $AdS_{3}$
with pure NS background. 
The latter formulation leads to the free field realization of the 
$sl(2|2)$ affine Lie superlalgebra. 
We have investigated the spacetime $N=4$ superconformal algebra in the 
light-cone gauge.
In the classical approximation,  we have obtained only half of the 
supercurrents in the background (\ref{eq:llg}).
In order to get the quantum superconformal algebra the BRST approach is
necessary\cite{SeWi}. 

As an application of the present formalism, it would be interesting to 
study the spacetime correlation functions of chiral primary fields.
In order to compute them, we need to investigate the bulk superstring 
theory\cite{deOoRoTa}, 
where we need to take into account the Liouville term in the action.
We may examine this GS action on $AdS_{3}$ as the WZW model on
the  $SU(2|2;{\bf C})/SU(2|2)$, which provides a spacetime 
supersymmetric extension
of the Euclidean $AdS_{3}=SU(2,{\bf C})/SU(2)$\cite{deOoRoTa}.
An element of the Lie supergroup $SU(2|2;{\bf C})$ admits the 
decomposition:
\begin{equation}
g=
\left(
\begin{array}{cccc}
1 & \G_{1} & \xi'_{12} & \xi'_{123} \\
0 & 1      & \xi_{2} & \xi_{23} \\
0 & 0      & 1       & \G_{3}  \\
0 & 0      & 0       &    1    \\
\end{array}
\right)
\left(
\begin{array}{cccc}
e^{\phi_{1}} & 0  & 0 & 0 \\
0 & e^{\phi_{2}}    & 0 & 0 \\
0 & 0      & e^{\phi_{3}}       & 0  \\
0 & 0      & 0       &   e^{\phi_{4}}    \\
\end{array}
\right)
h, \quad (h\in SU(2|2))
\end{equation}
where we put 
$
\xi'_{12}=\xi_{12}+\G_{1}\xi_{2}, \quad 
\xi'_{123}=\xi_{123}+\G_{1}\xi_{23}.
$
The Cartan part $\phi_{i}$ satisfy 
$\phi_{1}+\phi_{2}=0$
and $\phi_{3}+\phi_{4}=0$.
Then $SU(2|2)$ acts on $SU(2|2;{\bf C})$ by $g\rightarrow g h'$
($h'\in SU(2|2)$). 
The action of  WZW model on $SU(2|2;{\bf C})/SU(2|2)$ is defined by
$S=S_{WZW}(g g^{\dagger})$, where
\begin{equation}
gg^{\dagger}=
\left(
\begin{array}{cccc}
1 & \G_{1} & \xi'_{12} & \xi'_{123} \\
0 & 1      & \xi_{2} & \xi_{23} \\
0 & 0      & 1       & \G_{3}  \\
0 & 0      & 0       &    1    \\
\end{array}
\right)
\left(
\begin{array}{cccc}
e^{\phi_{1}} & 0  & 0 & 0 \\
0 & e^{\phi_{2}}    & 0 & 0 \\
0 & 0      & e^{\phi_{3}}       & 0  \\
0 & 0      & 0       &   e^{\phi_{4}}    \\
\end{array}
\right)
\left(
\begin{array}{cccc}
1 & 0 & 0 & 0 \\
\bar{\G}_{1} & 1      & 0 & 0 \\
\bar{\xi'}_{12} & \bar{\xi}_{2}      & 1       & 0  \\
\bar{\xi'}_{123} & \bar{\xi}_{23}      & \bar{\gamma}_{3}       &    1    \\
\end{array}
\right).
\end{equation}
Then the action becomes
\begin{eqnarray}
S&=&{1\over 2\pi} 
\int d^{2} z \biggl\{ -2 \pa\phi_{1}\bar{\pa}\phi_{1}
+2 \pa\phi_{3}\bar{\pa}\phi_{3}
+\pa\bar{\gamma}_{1}\bar{\pa}\gamma_{1} e^{\phi_{1}-\phi_{2}}
+\pa\bar{\gamma}_{3}\bar{\pa}\gamma_{3} e^{\phi_{3}-\phi_{4}}
+e^{\phi_{2}-\phi_{3}} \pa\bar{\xi}_{2}\bar{\pa}\xi_{2}  \CR
&& 
+e^{\phi_{1}-\phi_{3}} (\pa\bar{\xi}_{12}+\bar{\gamma}_{1}\pa\bar{\xi}_{2}) 
                       (\bar{\pa}\xi_{12}+\G_{1}\bar{\pa}\xi_{2})
+e^{\phi_{2}-\phi_{4}} (\pa\bar{\xi}_{23}-\bar{\gamma}_{3}\pa\bar{\xi}_{2}) 
                       (\bar{\pa}\xi_{23}-\gamma_{3}\bar{\pa}\xi_{2}) \CR
&& \!\!\!
+e^{\phi_{1}-\phi_{4}}
(\pa\bar{\xi}_{123}-\bar{\gamma}_{3}\pa\bar{\xi}_{12}
+\bar{\gamma}_{1}\pa\bar{\xi}_{23} 
-\bar{\gamma}_{1}\bar{\gamma}_{3}\pa\bar{\xi}_{2}) 
(\bar{\pa}\xi_{123}-\G_{3}\bar{\pa}\xi_{12}
+\G_{1}\bar{\pa}\xi_{23} 
-\G_{1}\G_{3}\bar{\pa}\xi_{2}) 
\biggr\}. \CR
\end{eqnarray}
Introducing the auxiliary fields $\eta_{i}$ conjugate to $\xi_{i}$ for
$i=2,12,23,123$ and $\B_{a}$ to $\G_{a}$ ($a=1,3$), we obtain
\begin{eqnarray}
S&=&{1\over 2\pi} 
\int d^{2} z \biggl\{ 
-2 \pa\phi_{1}\bar{\pa}\phi_{1}
+2 \pa\phi_{3}\bar{\pa}\phi_{3}
+\sum_{a}(\B_{a}\bar{\pa}\G_{a}+\bar{\beta}_{a}\partial\bar{\gamma}_{a})
+\sum_{i}(\eta_{i}\bar{\pa}\xi_{i}+\bar{\eta}_{i}\bar{\pa}\bar{\xi}_{i})
 \CR
&&
 -\B_{1}\bar{\B}_{1} e^{\phi_{2}-\phi_{1}}
-\B_{3}\bar{\B}_{3} e^{\phi_{4}-\phi_{3}}
-e^{\phi_{4}-\phi_{1}} \bar{\eta}_{123}\eta_{123}
-e^{\phi_{3}-\phi_{1}}
(\bar{\eta}_{12}+\bar{\gamma}_{3}\bar{\eta}_{123}) 
 (\eta_{12}+\gamma_{3}\eta_{123}) 
\CR
&& 
-e^{\phi_{3}-\phi_{2}}
(\bar{\eta}_{2}-\bar{\gamma}_{1}\bar{\eta}_{12}
+\bar{\gamma}_{3}\bar{\eta}_{23}
-\bar{\gamma}_{1}\bar{\gamma}_{3}\bar{\eta}_{123})
(\eta_{2}-\gamma_{1}\eta_{12}+\gamma_{3}\eta_{23}
-\gamma_{1}\gamma_{2}\eta_{123})
\CR
&& -e^{\phi_{4}-\phi_{2}} 
(\bar{\eta}_{23}-\bar{\gamma}_{1}\bar{\eta}_{123}) 
(\eta_{23}-\gamma_{1}\eta_{123})
\biggr\}.
\end{eqnarray}
In order to obtain the free field realization (\ref{eq:waki}), we find
that $i\alpha_{1}\cdot\varphi=\phi_{1}-\phi_{2}$, 
$i\alpha_{3}\cdot\varphi=\phi_{3}-\phi_{4}$,
$(\tilde{\beta}_{\alpha_{a}},\tilde{\gamma}_{\alpha_{a}})
=(\beta_{a},\gamma_{a})$ ($a=1,3$) and 
$(\tilde{\eta}_{\alpha_{2}},\tilde{\xi}_{\alpha_{2}})=
(\bar{\eta}_{2}-\bar{\gamma}_{1}\bar{\eta}_{12}
+\bar{\gamma}_{3}\bar{\eta}_{23}
-\bar{\gamma}_{1}\bar{\gamma}_{3}\bar{\eta}_{123},\xi_{2})$,
$(\tilde{\eta}_{\alpha_{1}+\alpha_{2}},\tilde{\xi}_{\alpha_{1}+\alpha_{2}})=
(\eta_{12}+\gamma_{3}\eta_{123},\xi_{12})
$,
$(\tilde{\eta}_{\alpha_{2}+\alpha_{3}},\tilde{\xi}_{\alpha_{2}+\alpha_{3}})=
(\eta_{23}-\gamma_{1}\eta_{123},\xi_{23})
$ and
$(\tilde{\eta}_{\alpha_{1}+\alpha_{2}+\alpha_{3}},
\tilde{\xi}_{\alpha_{1}+\alpha_{2}+\alpha_{3}})=
(\eta_{123},\xi_{123})
$.
If we consider the quantum correction, the vertex operators 
$\B_{1}\bar{\B}_{1}
e^{\phi_{2}-\phi_{1}}$, $\B_{3}\bar{\B}_{3} e^{\phi_{4}-\phi_{3}}$ and 
$e^{\phi_{2}-\phi_{3}}
(\eta_{2}-\gamma_{1}\eta_{12}+\gamma_{3}\eta_{23}
-\gamma_{1}\gamma_{3}\eta_{123})
(\bar{\eta}_{2}-\bar{\gamma}_{1}\bar{\eta}_{12}
+\bar{\gamma}_{3}\bar{\eta}_{23}
-\bar{\gamma}_{1}\bar{\gamma}_{3}\bar{\eta}_{123})$
become the screening operators, which commutes with the 
$sl(2|2)^{(1)}$ currents.
The remaining vertex operators, however, do not commute with the
currents corresponding to the positive roots\cite{deSh}.
It is not clear at present how these non-screening operators work 
for studying the spacetime correlation functions in the present GS formalism. 

It is also interesting to generalize the present construction to 
other models with various spacetime supersymmetries.
For example, we may study the GS superstrings on 
$AdS_{3}\times S^{3}\times S^{3}\times S^{1}$.
The NSR formulation is discussed in \cite{ElFeGiTs}.
In this case, we expect the $D(2|1;\alpha)^{(1)}$ affine Lie superalgebra
to represent spacetime supersymmetry. 
The light-cone gauge choice would lead to the linear subalgebra of the 
$N=4$ $SU(2)\times SU(2)$ non-linearly extended superconformal algebra.
Another interesting example is the GS superstrings with $N<4$ 
spacetime supersymmetries\cite{GR}. 
For $N=2$ case, we may study
$AdS_{3}\times S^{1}\times {\cal N}/U(1)$, where ${\cal N}/U(1)$ denotes
certain $N=2$ superconformal Kazama-Suzuki model. 
{}From this model, we obtain the non-critical superstrings with 
$N=2$ spacetime superconformal symmetry\cite{GiKuPe}.
In this case, we expect that 
the affine Lie superalgebra $sl(2|1)^{(1)}$ \cite{It} 
provides the corresponding GS superstrings.
Since one can introduce non-trivial RR backgrounds in this
formulation\cite{BeVaWi}, it is possible to study the singular 
CFT resolved by non-trivial background \cite{GuVaWi}.
These subjects will be discussed elsewhere.

\vskip3mm\noindent
This work is supported in part by 
the Grant-in-Aid from the Ministry of Education, Science and Culture,
Priority Area: \lq\lq Supersymmetry and Unified
Theory of Elementary Particles'' (\#707).


\end{document}